\documentclass[twocolumn,twoside]{article}
\usepackage{graphics}
\newcommand{\hb}{\\ \hspace*{2ex}}
\newcommand{\hc}{\\ \hspace*{3ex}}
\begin{document}
\title{X-RAY EMISSION AND ORIENTATION \\OF SELECTED PF GALAXY CLUSTERS}
\author{A.V.\,Tugay$^{1}$, S.S.\,Dylda$^{1}$, E.A.\,Panko$^{2}$\\[2mm] 
\begin{tabular}{l}
 $^1$  Astronomy and Space Physics Department, Faculty of Physics, \\ Taras Shevchenko National University of Kyiv,\hb
 Glushkova ave., 4, Kyiv, 03127, Ukraine,  {\em tugay.anatoliy@gmail.com}\\
$^2$ Department of Theoretical Physics and Astronomy,\\ I.I.\,I. I. Mechnikov Odessa National University,\hb
 Shevchenko Park, Odessa, 65014, Ukraine,  {\em panko.elena@gmail.com}\\[2mm]
\end{tabular}
}
\date{}
\maketitle

ABSTRACT.
X-ray counterparts for 35 galaxy clusters contained in the PF catalog of galaxy clusters and groups were found in  XMM-Newton archive. 22 ones (all from ACO catalogue) have extended elliptic X-ray haloes appropriate for determination of orientation. Position angles and eccentricities were calculated and compared with cluster orientations optical band.

\\[3mm]

{\bf 1. Introduction}\\[1mm]

The study of morphology of galaxy clusters is important for understanding the large scale structure of Universe.
Orientation of galaxies and clusters may give information about clusterisation and cosmologic evolution. The best way to consider orientation of extragalactic objects is the usage of special large and uniform catalog of galaxy clusters. Clusters are also suitable for orientation analysis in X-ray band because they contain a haloes of hot intergalactic gas.
\begin{table*}
\begin{center}
\caption{Orientation and eccentricitiesof X-ray PF galaxy clusters.}
\begin{tabular}{crclcc}
\hline
 PF        & ACO  & PA  & $ PA_X $    & $ e_{PF} $ & $ e_X $ \\
\hline
 0004-3606 & 2717 & 160 &  36$\pm $17 & 0.29 & 0.37$\pm $0.08 \\
 0009-3469 & 2721 &  93 &  12$\pm $6  & 0.23 & 0.67$\pm $0.02 \\
 0022-1954 &   13 & 102 &  57$\pm $3  & 0.13 & 0.64$\pm $0.04 \\
 0034-2570 &   22 & 174 & 105$\pm $7  & 0.13 & 0.58$\pm $0.17 \\
 0042-3308 & S 41 & 163 & 118$\pm $1  & 0.14 & 0.60$\pm $0.01 \\
\hline
 0068-2875 & 2811 & 119 &  65$\pm $10 & 0.24 & 0.48$\pm $0.05 \\
 0082-2951 & S 84 &   6 &  58$\pm $11 & 0.17 & 0.48$\pm $0.06 \\
 0104-2195 &  133 &  65 &  16$\pm $6  & 0.09 & 0.55$\pm $0.03 \\
 0115-4600 & 2877 & 179 &  41$\pm $29 & 0.17 & 0.29$\pm $0.09 \\
 0168-5458 & 2933 &  70 &  22$\pm $3  & 0.08 & 0.75$\pm $0.30 \\
\hline
 0229-4765 &S 239 & 148 & 198$\pm $8  & 0.16 & 0.59$\pm $0.06 \\
 0329-4427 & 3112 &  17 &  12$\pm $1  & 0.14 & 0.63$\pm $0.02 \\
 0350-5258 & 3128 & 162 & 139$\pm $5  & 0.14 & 0.91$\pm $0.05 \\
 0370-5364 & 3158 &  56 &  11$\pm $6  & 0.11 & 0.50$\pm $0.05 \\
 0480-3720 &  514 &  61 &  35$\pm $1  & 0.20 & 0.67$\pm $0.01 \\
\hline
 0500-3868 & 3301 &  43 &  69$\pm $3  & 0.08 & 0.48$\pm $0.02 \\
 2020-5671 & 3667 &  39 &  31$\pm $5  & 0.13 & 0.74$\pm $0.06 \\
 2070-3523 & 3705 & 122 & 115$\pm $6  & 0.29 & 0.47$\pm $0.04 \\
 2149-5088 & 3771 &   0 & 148$\pm $58 & 0.13 & 0.62$\pm $0.19 \\
 2181-3068 & 3814 &  60 &  16$\pm $4  & 0.17 & 0.50$\pm $0.03 \\
\hline
 2187-1958 & 2384 & 167 & 104$\pm $5  & 0.29 & 0.81$\pm $0.05 \\
 2229-3570 & 3854 &  27 &  57$\pm $10 & 0.29 & 0.41$\pm $0.10 \\
\hline
\end{tabular}
\end{center}
\end{table*}

Since (Binggeli, 1984) orientation of galaxies in clusters was the subject of numerous studies.
Orientations of galaxies in 247 rich Abell clusters were studied in detail in Godlowski et al. (2010) and Panko et al. (2013) with corresponding statistical analysis and simulations. Orientation of galaxies from compact sample can be numerically described by the distribution of anisotropy parameter. The parameter was calculated for edge-on galaxies in Parnovsky \& Tugay (2007) and for nearby galaxy groups in Godlowski et al.(2012). Orientation of galaxies in nearby groups was studied by Pajowska et al.(2012) too. \\ [2mm]

\pagebreak

{\bf 2. Observational data}\\[1mm]

Our study in optic band based on PF catalogue of galaxy clusters and groups data. The catalogue contains 6188 structures of southern sky (Panko \& Flin, 2006). Orientations and shapes of PF clusters were calculated taking into consideration galaxy $2D$ locations in the cluster field using the  covariance ellipse method (Carter \& Metcalfe, 1980; Biernacka et al., 2007).
To select PF clusters counterparts in X-rays we used Xgal list of all X-ray extragalactic sources observed by XMM-Newton space observatory (Tugay, 2012).
Xgal includes 5021 sources and approximately 30\% of them are galaxy clusters.
In the current study we found PF clusters counterparts in X-rays, calculated their orientations and eccentricities in X-ray band and compared obtained values with optical data.\\[6mm]

{\bf 3. Method}\\[1mm]

Cross-correlation of PF and Xgal objects was performed on the base of condition of appearing Xgal source within PF cluster radius. We found 35 Xgal sources counterparts in PF catalogue.
To estimate orientation of X-ray halo we selected at XMM images pixels with certain numbers of detected photons (two, three, four etc). Then we approximated each set of pixels with ellipse by the minimal square method and found positional angle PA and eccentricity {\it e}.
We succeeded to find X-ray orientation for 28 Abell clusters (Abell, Corwin \& Olowin, 1989) from PF catalog but 6 of them have no appropriate optical orientation. Common results are presented in Table 1. Table 2 shows PF clusters with X-ray sources that are unappropriate for orientation determination by any reason.\\[2mm]

{\bf 3. The general notes on selected clusters }\\[1mm]

Here are general notes on selected clusters.

A\,2717, A\,2877, A\,S\,1111. These clusters looks like spherical, but for first and second ones we determined PA.

A\,13, A\,2811, AO\,S\,84, A\,S\,239, AO\,3158, A\,3667, A\,3771, A\,3856. Orientation of cluster core differs from periphery. We didn't determine PA for A\,3856.

Double clusters: A\,2933, A\,3128, A\,2384.

Faint clusters: A\,S\,41, A\,514, A\,3301.

A\,3705 - a pair of interacting clusters, but PA was found by X-ray image.

Clusters with undefined orientation or with large differences in optical and X-ray PA are presented in Table 3.

\begin{table}[h]
\caption{PF clusters with X-ray sources not suitable for comparison of orientation.}
\begin{tabular}{lrclc}
\hline
 PF        & ACO  & PA  & $ PA_X $    & note \\
\hline
 0093-2244 & MCXC   & 131 & -           & point source\\
 0120-3828 & BAX    & 106 & -           & point source\\
 0263-5237 & 3038   & 146 & -           & point source\\
 0408-3720 & new    & 180 & -           & point source\\
 0451-6138 & 3266   & 106 & -           & interacting \\
\hline
 0532-2498 & Snow 20&  24 & -           & point source\\
 2230-3890 & 3856   & -   &  50$\pm $4  & diff. orient.\\
 2256-3778 & 3888   & -   & 166$\pm $6  & \\
 2277-5266 & 3911   & -   &  50$\pm $4  & \\
 2323-4265 & S 1101 & -   & 147$\pm $1  & \\
\hline
 2331-4225 & S 1111 & -   & -           & spherical \\
 2378-2816 & 4038   & -   &  32$\pm $6  & \\
 2395-3453 & 4059   & -   &  58$\pm $5  & \\
\hline
\end{tabular}
\end{table}


Except for 22 clusters with both optical and X-ray orientation there are 13 PF clusters with X-ray sources for which it is impossible to compare orientations.

4 point X-ray sources were found within PF clusters. No X-ray haloes of galaxy clusters were detected for these systems.

1. PF\,0120-3828. BAX 017.9025-38.1867 cluster.
X-ray source  2MASXJ00570192-3806028 galaxy (only 3 references in SIMBAD).

2. PF\,0093-2244. MCXC J0056.9-2213 cluster.
X-ray source  RBS139 Seyfert 1 galaxy.

3. PF\,0263-5237. Abell 3038 cluster.
X-ray source  ESO 198-24 Seyfert 1 galaxy.

4. PF\,0532-2498. Snow 20 cluster (T.Snow, 1970).
X-ray source - IC\,411 galaxy (9 references).

PF\,0408-3720 cluster was unknown in previous works. In the region of this cluster we found ESO 359-19 Seyfert 1 galaxy.

A\,3266 - complex system of interacting clusters unappropriate for orientation estimation in X-rays.

Seven PF clusters (bottom of Table 2) does not show anisotropy in optical band, so we excluded them from comparison with X-ray data. Specifically, A\,3856 shows different orientation of the core and periphery of X-ray halo; A\,1111 has spherical halo; other clusters include A\,1101, A\,3888, A\,3911, A\,4038 and A\,4059. \\[2mm]

{\bf 4. Results and conclution}\\[1mm]

Analysis of Table 1 shows that PA tend to correlate. Eccentricity is larger in X-rays because visible hot gas halo lies close to the center of cluster in the region of larger potential. The correlation of orientation in two bands leads to issue that galaxies and gas halo in clusters are involved in significant gravitational interaction but the processes of cluster evolution continue in the current cosmological era. \\[2mm]


{\it Acknowledgements.} This research has made use of NASA's
Astrophysics Data System. \\[1mm]

{\bf References\\[1mm]}
\noindent\\
Abell G.O., Corwin H.G., Olowin, R.P.: 1989, {\it ApJS}, {\bf 70}, 1.\\
Biernacka M., Flin P., Panko E., Juszczyk T.: 2007,  {\it Odessa Astron. Publ.}, {\bf 20}, 26.\\
Binggeli B.: 1984, {\it A\& A}, {\bf 107}, 338 \\
Carter D., Metcalf, N.: 1980 {\it MNRAS}, {\bf191}, 325.
Godlowski W., Piwowarska P., Panko E., Flin P.: 2010, {\it ApJ}, {\bf 723}, 985.\\
Godlowski M., Panko E., Pajowska P., Flin P.: 2012, {\it JPhSt.}, {\bf 16}, 3901.\\
Pajowska P., Godlowski M., Panko E., Flin P.: 2012, {\it JPhSt.}, {\bf 16}, 4901.\\
Panko E., Flin P.: 2006, {\it JAD.,} {\bf 12}, 1.\\
Panko E., Piwowarska P., Godlowska J., Godlowski W., Flin P.: 2013, {\it Ap.}, {\bf 56}, 322.\\
Parnovsky S., Tugay A.: 2007, {\it JPhSt.}, {\bf 11}, 366.\\
Tugay A.V.: 2012, {\it Odessa Astron. Publ.}, {\bf 25}, 142. \hc {\it http://arxiv.org/abs/1311.4333} \\

\vfill

\newpage

\mbox{}

\newpage

\begin{table}[h]
\caption{Notes on large differences in orientation. Most of such clusters has near-spherical X-ray halo. Optical orientation of last 5 clusters can not be determined because of their diffuse structure.}
\begin{tabular}{lrrrr}
\hline
 Name                                 &        RA         &       DEC      &     PA                  &  e                 \\
\hline
ACO 2700                              &    0.9567         &    2.0649      &     25  $\pm $   11     &  0.51 $\pm $    0.2\\         
ACO 119                               &   14.0675         &   -1.2489      &    110  $\pm $    3     &  0.09 $\pm $   0.01\\         
ACO 122                               &   14.3449         &   -26.281      &     84  $\pm $    2     & 0.713 $\pm $  0.327\\         
ACO 2984                              &   32.8544         &   -40.291      &     84  $\pm $    1     &  0.52 $\pm $    0.4\\         
ACO 399                               &   44.4756         &   13.0317      &     63  $\pm $   15     &  0.31 $\pm $    0.1\\         
ACO 401                               &   44.7414         &   13.5821      &     53  $\pm $    1     &  0.39 $\pm $   0.25\\         
ACO 3112                              &   49.4902         &  -44.2385      &     71  $\pm $    5     &  0.62 $\pm $    0.5\\         
ACO 3158                              &   55.7197         &  -53.6286      &     82  $\pm $    4     & 0.212 $\pm $   0.01\\         
ACO S 384                             &   56.4429         &   -41.204      &     38  $\pm $   20     &  0.36 $\pm $   0.25\\         
ClG 0422-09                           &   66.4638         &   -8.5605      &     19  $\pm $    1     & 0.079 $\pm $   0.01\\         
ACO 496                               &   68.4069         &  -13.2603      &      8  $\pm $    3     &   0.2 $\pm $    0.1\\         
ClG 0451-03                           &   73.5469         &   -3.0146      &     71  $\pm $    8     &  0.17 $\pm $   0.02\\         
MCXC J0528.9-3927                     &   82.2195         &  -39.4722      &     46  $\pm $   39     &  0.28 $\pm $   0.16\\         
MCXC J0532.9-3701                     &   83.2326         &  -37.0268      &     88  $\pm $    1     &  0.19 $\pm $   0.14\\         
ACO 3378                              &   91.4753         &  -35.3023      &    111  $\pm $   20     &  0.45 $\pm $   0.21\\         
ACO 3391                              &   96.5869         &  -53.6921      &    107  $\pm $    3     &  0.22 $\pm $   0.04\\         
ACO 3404                              &   101.371         &  -54.2267      &     45  $\pm $    2     &  0.47 $\pm $    0.3\\         
ZwCl 0735+7421                        &   115.435         &   74.2439      &     42  $\pm $   11     &  0.15 $\pm $   0.08\\         
ClG 0745-1910                         &   116.881         &  -19.2952      &     76  $\pm $    2     &  0.17 $\pm $   0.46\\         
ACO 653                               &    125.46         &    1.2003      &         $\pm $          &       $\pm $       \\         
ACO 689                               &   129.353         &   14.9722      &     46  $\pm $   34     &  0.14 $\pm $   0.01\\         
ACO 773                               &    139.47         &   51.7275      &    103  $\pm $    5     &  0.37 $\pm $   0.25\\         
ACO 901A                              &   149.117         &   -9.9554      &     41  $\pm $   20     &  0.36 $\pm $   0.27\\         
ACO 907                               &   149.592         &  -11.0637      &     45  $\pm $   22     &  0.42 $\pm $    0.3\\         
ZwCl 1021+0426                        &   155.916         &    4.1864      &     35  $\pm $    9     &  0.46 $\pm $    0.4\\         
ACO 1084                              &   161.137         &   -7.0688      &     20  $\pm $    6     &  0.51 $\pm $   0.33\\         
ACO 1201                              &   168.228         &   13.4329      &     25  $\pm $   15     &   0.5 $\pm $    0.2\\         
ClG J1115+5319                        &   168.811         &   53.3323      &         $\pm $          &       $\pm $       \\         
ACO 1413                              &   178.825         &    23.405      &      4  $\pm $    2     &  0.63 $\pm $   0.35\\         
MCXC J1206.2-0848                     &   181.552         &   -8.8018      &     59  $\pm $   25     &  0.24 $\pm $   0.17\\         
ZwCl 1215+0400                        &    184.42         &    3.6574      &    144  $\pm $    9     &  0.49 $\pm $   0.15\\         
ACO S 700                             &   189.172         &  -33.9246      &      8  $\pm $    3     &  0.07 $\pm $   0.01\\         
ACO 3528                              &   193.593         &   -29.013      &     10  $\pm $    1     &   0.6 $\pm $    0.4\\         
ACO 1651                              &   194.844         &   -4.1954      &     99  $\pm $    3     &  0.35 $\pm $   0.27\\         
ACO 1656                              &   194.944         &   27.9699      &     73  $\pm $    3     &  0.18 $\pm $    0.1\\         
ACO 1663                              &   195.718         &   -2.5148      &     48  $\pm $   39     &   0.3 $\pm $    0.2\\         
ACO 1664                              &   195.927         &  -24.2452      &    157  $\pm $    5     &  0.48 $\pm $   0.25\\         
2E 2975                               &   197.329         &   -1.6224      &     60  $\pm $   15     &  0.28 $\pm $   0.25\\         
1RXS J132441.9-573650                 &   201.195         &  -57.6103      &     45  $\pm $   33     &  0.17 $\pm $    0.1\\         
ACO 3558                              &   201.985         &   -31.495      &     43  $\pm $   16     &  0.47 $\pm $    0.3\\         
ACO 1750N                             &   202.795         &   -1.7292      &    140  $\pm $   25     & 0.379 $\pm $   0.24\\         
ACO 3560                              &   203.112         &  -33.1415      &     26  $\pm $   14     &  0.32 $\pm $   0.27\\         
ACO 3562                              &   203.397         &  -31.6744      &    150  $\pm $   29     &  0.31 $\pm $   0.14\\         
ACO 1775                              &   205.456         &   26.3719      &    136  $\pm $   40     &  0.31 $\pm $   0.23\\         
ACO 3571                              &   206.869         &  -32.8622      &     84  $\pm $    2     &   0.3 $\pm $   0.01\\         
ClG J1347-1145                        &   206.879         &  -11.7532      &     11  $\pm $    6     &  0.15 $\pm $   0.08\\         
ACO 1835                              &   210.258         &    2.8783      &     10  $\pm $    5     &  0.21 $\pm $   0.16\\         
ACO 3581                              &   211.871         &  -27.0173      &     69  $\pm $    7     &   0.2 $\pm $   0.06\\         
2XMM J141830.6+251052                 &   214.628         &   25.1812      &     30  $\pm $    3     &  0.45 $\pm $    0.3\\         
NGC 5718 Group                        &   220.175         &    3.4679      &     61  $\pm $    2     &  0.22 $\pm $   0.15\\         
ClG J1504-0248                        &   226.032         &   -2.8046      &    147  $\pm $   17     &   0.1 $\pm $   0.06\\         
ACO 2050                              &   229.076         &    0.0919      &    141  $\pm $   34     &  0.47 $\pm $   0.25\\         
ACO 2052                              &   229.172         &    7.0026      &    135  $\pm $    6     &  0.48 $\pm $    0.3\\         
ACO 2051                              &   229.185         &   -0.9698      &     64  $\pm $    8     &       $\pm $       \\         
ACO 2055                              &   229.691         &    6.2322      &     12  $\pm $   12     &  0.31 $\pm $   0.13\\         
ACO 2063                              &    230.77         &    8.6079      &      8  $\pm $    4     &  0.25 $\pm $   0.06\\         
ACO 2204                              &   248.196         &    5.5755      &      7  $\pm $    5     & 0.087 $\pm $   0.01\\         
ClG J1720+2638                        &   260.042         &   26.6252      &     70  $\pm $    3     &  0.38 $\pm $   0.01\\         
MCXC J2011.3-5725                     &   302.863         &  -57.4193      &     60  $\pm $   15     &  0.28 $\pm $   0.25\\         
ACO 3667                              &   303.112         &  -56.8345      &      8  $\pm $    8     &  0.76 $\pm $   0.08\\         
2MAXI J2014-244                       &   303.715         &  -24.5059      &      9  $\pm $    1     &  0.52 $\pm $   0.45\\         
ACO 3693                              &   308.583         &  -34.4935      &     39  $\pm $   23     &  0.12 $\pm $   0.08\\         
ClG J2129+0005                        &   322.416         &    0.0891      &     24  $\pm $    1     &   0.5 $\pm $    0.3\\         
ACO 3814                              &   327.281         &  -30.7011      &     65  $\pm $   23     &  0.56 $\pm $   0.27\\         
XMMXCS J221656.6-172527.2             &   334.236         &  -17.4235      &     71  $\pm $    6     &  0.22 $\pm $    0.2\\         
ACO 3854                              &   334.441         &  -35.7253      &     37  $\pm $   22     &  0.51 $\pm $   0.37\\         
ACO 3856                              &   334.667         &  -38.9008      &     39  $\pm $   10     &  0.56 $\pm $   0.31\\         
ACO S 1101                            &   348.494         &  -42.7272      &    121  $\pm $    2     &  0.45 $\pm $    0.3\\         
ACO 3992                              &   349.919         &  -73.2259      &      2  $\pm $    2     &  0.37 $\pm $    0.1\\         
ACO 2597                              &   351.332         &  -12.1243      &     37  $\pm $    1     &  0.17 $\pm $   0.05\\         
ACO 4010                              &   352.807         &  -36.5111      &    148  $\pm $    4     & 0.148 $\pm $   0.01\\         
ACO 2626                              &   354.127         &   21.1462      &    156  $\pm $    5     &  0.67 $\pm $   0.06\\         
ACO 2667                              &   357.914         &  -26.0841      &    133  $\pm $   27     & 0.403 $\pm $   0.28\\         
ACO 2670                              &   358.562         &  -10.4172      &    132  $\pm $   35     &  0.23 $\pm $   0.14\\         
ACO 13                                &    3.4044         &  -19.4976      &     60  $\pm $   20     &  0.41 $\pm $    0.1\\         
2XMMi J001737.3-005239                &    4.4058         &   -0.8777      &    149  $\pm $    2     &       $\pm $       \\         
ClG 0016+16                           &    4.6394         &   16.4368      &         $\pm $          &       $\pm $       \\         
\hline
\end{tabular}
\end{table}

\end{document}